\documentclass[twocolumn,usegraphicx,usenatbib,useAMS]{mn2e}
\newcommand{\nit}{\noindent}
\newcommand{\ud}{\mathrm{d}}

\newcommand{\beq}{\begin{equation}}
\newcommand{\eeq}{\end{equation}}
\newcommand{\ba}{\begin{eqnarray}}
\newcommand{\ea}{\end{eqnarray}}
\newcommand{\etal}{et al.\ }

\def\gtsima{$\; \buildrel > \over \sim \;$}
\def\ltsima{$\; \buildrel < \over \sim \;$}
\def\gsim{\lower.5ex\hbox{\gtsima}}
\def\lsim{\lower.5ex\hbox{\ltsima}}
\def\msun{{M_\odot}}

\begin{document}
\title[Dwarf Galaxy Suppression]{Suppression of Dwarf Galaxy Formation
by Cosmic Shocks} \author[Filippo Sigward et al.]{Filippo Sigward,$^1$
Andrea Ferrara$^2$ \& Evan Scannapieco$^3$ \\ $^1$Universit\`a di
Firenze, Dipartimento di Astronomia e Scienza dello Spazio, Largo
Enrico Fermi 5, 50125 Firenze, Italy \\ $^2$SISSA/International School
for Advanced Studies, Via Beirut 4, 34014 Trieste, Italy \\ $^3$Kavli
Institute of Theoretical Physics, University of California, Santa
Barbara, CA 93106, USA}
%\email{sigward@arcetri.astro.it}

\maketitle

\begin{abstract}

We carry out a numerical study of the effects of supernova-driven
shocks on galaxy formation at $z=9$. These ``cosmic explosions'' can
have a significant impact on galaxies forming nearby. We study such
interactions in two key cases. In the first case in which the forming
galaxy has already virialized, the impinging shock has only a small
effect ($\lsim 1\%$ of the gas is removed) and star formation continues
relatively unimpeded. However, in the second case in which the nearby
forming galaxy is at the more diffuse turn-around stage, a large
fraction ($\approx 70\%$) of the gas is stripped away from the host
dark-matter halo and ejected into the intergalactic medium. As the
time spent near turn-around is much longer than the interval from
virialization to galaxy formation due to strong radiative losses, we
expect the second case to be more representative of the majority of
outflow-galaxy interactions. Thus SN-driven pregalactic outflows may
be an efficient mechanism for inhibiting the formation of neighbouring
galaxies at high redshift. We briefly outline the possible
cosmological consequences of this effect.
\end{abstract}


\section{Introduction}

In currently favoured cosmological scenarios, the formation of structure is a hierarchical
process, in which small initial subunits merge and accrete diffuse material to form larger
structures of ever increasing mass.  For instance, in the popular Cold Dark Matter (hereafter CDM)
cosmogony\footnote{Throughout this paper we will assume a flat universe with total matter, vacuum,
and baryonic densities in units of the critical density of $\Omega_m = 0.3$, $\Omega_{\Lambda} =
0.7$, and $\Omega_b = 0.05$ respectively, and a Hubble constant of $H_0 = 100\,h$ km s$^{-1}$
Mpc$^{-1}$, with $h= 0.65$.}, large numbers of objects with masses $M=10^8 h^{-1} M_\odot$
collapse at $z \sim 9$ from 2--$\sigma$ density fluctuations. In this collapse, the gas first
falls inwards along with the dark matter, then shock-heats to the virial temperature ($T_{\rm
vir}  \simeq 10^4$ K; see, however, Kere\v{s} \etal 2004 who find that for low-mass objects a cold
accretion mode can take place), condenses rapidly due to line cooling, and  becomes
self-gravitating. Subsequently, massive stars form, synthesize heavy elements, and explode as
supernovae (SNe) after $\sim 10^7$ yr, enriching the ambient medium. These processes may have been
enhanced by primordial zero-metallicity stars (Population III stars) that may have been massive
enough to ignite powerful pair-production SNe (e.g. \cite{HW02}), or generate large `seed' black
holes \citep{MR01}.

Around a redshift of 10-15, early sub-galactic stellar systems,
perhaps aided by a population of accreting black holes in their
nuclei, generated the ultraviolet radiation and mechanical energy that
reheated and reionized most of the hydrogen in the universe.  The
latest analysis of the data from the WMAP (\emph{Wilkinson Microwave
Anisotropy Probe}) satellite seems to suggest that this reionization
took place at $z_{\rm ion} = 20^{+10}_{-9}$ \citep{Koetal03},
requiring the production of ionizing photons to have been extremely
efficient in high-redshift galaxies (e.g.  \cite{CFW03}).

The history of this crucial stage in cosmic structure formation
depends both on the power spectrum of small-scale density fluctuations
and on a complex network of poorly understood ``feedback''
processes. A generic consequence of all hierarchical scenarios is a
large deposition of energy by SNe in the shallow potential wells of
early sub-galactic systems.  This may have two main effects, depending
on the efficiency with which halo baryons can cool and fragment into
clouds and then into massive stars: (a) the disruption of the host
(dwarf) galaxy -- the most violent version of stellar feedback --
\citep{La74,DS86,MacLF99}; (b) the ejection of metal-enriched gas from
the host galaxy and the subsequent early pollution of the
intergalactic medium (IGM)
\citep{TSE93,CO99,SB01,MFR01,SFM02,Sp03}. Furthermore, the
well-established presence of heavy elements such as carbon, nitrogen
and silicon in the Ly$\alpha$ forest clouds at $z = 3-3.5$ provides
strong evidence for such an early episode of pregalactic star
formation and outflows \citep{So01, PP04}.

Similarly, stellar feedback and pregalactic outflows are likely to
have efficiently inhibited high-redshift galaxy (and consequently
star) formation.  Because of the short cooling time of the gas in
high-$z$ small halos, strong feedback has been advocated in
hierarchical clustering scenarios to prevent a ``cooling catastrophe''
in which too many baryons are converted into stars at very high
redshifts. The required reduction of the stellar birthrate in halos
with low escape velocities may naturally result from the heating and
expulsion of material by winds from massive stars and quasars, and
multiple SN explosions.

It is also well known that the radiative and mechanical energy
deposited by massive stars and accreting black holes into the
interstellar medium of protogalaxies may have a more global negative
feedback on galaxy formation. The photoionizing background responsible
for the reionization of the IGM will have two effects: a gas pressure
increase, preventing accretion into low-mass halos, and the reduction
of the rate of radiative cooling of baryons inside the halos
themselves
\citep{Ef92,TW96,NS97,Beetal01,Beetal02a,Beetal02b,Beetal03}. Furthermore,
blast-waves produced by mini-quasars and protogalaxies propagate into
the intergalactic space and may drive large portions of the IGM to a
much higher adiabat than expected from photoionization
\citep{Vo96,MFR01, TMS01,CB01}, so as to quench the collapse of
further galactic systems by raising the cosmological Jeans mass.

In a complementary study, \cite{BM03} have recently performed a
detailed calculation of the effect of the increased gas pressure after
this global early energy input into the IGM at the end of the cosmic
dark ages. They have shown that preheating alone is not able to
explain the sharp cut-off of the luminosity function at bright
magnitudes and reproduce the observed abundance of satellite galaxies
in the Local Group. It is worth noting that preheating consists of a
homogeneous heat deposition in the IGM. Therefore, it affects galaxy
formation in a global manner, through an increase in the filtering
mass \citep{Gn00}.

In this paper we will be concerned with events of multiple supernovae
in protogalaxies and the mechanical interaction of their shocks with
neighbouring overdense regions.  Considerable recent theoretical
effort has gone into understanding the long-range feedback due to
UV/ionizing radiation. Relatively less attention has been paid to the
problem of how galaxy formation in these objects is hindered by shocks
from nearby protogalaxies. However, limited analytic and numerical
studies on a global cosmological scale have indicated that such
interactions may have had an enormous impact on the galaxy luminosity
function and its evolution \citep{SFM02,TSD02}. While shock-cloud
interactions within the interstellar medium have been simulated in
detail (e.g.\ \cite{KMC94}), such related cosmological interactions
have been largely neglected. Thus it is important to study this
process in detail.

The structure of this paper is as follows: in \S\ref{model} we outline
the model used and the assumptions adopted in building up our model of
``cosmic" stellar feedback; in \S\ref{simul} we describe the
simulations carried out to examine the effects of SN explosions on a
nearby halo, and in \S\ref{results} we present our results and give a
quantitative analysis.  A summary is given in \S\ref{summary}.

\section{Model assumptions}
\label{model}

We want to study the effect of cosmic explosions produced by SNe in a
given (source) halo impinging onto a neighbouring (target) object on
its way to collapse into a luminous galaxy. To address this problem in
a numerical context, we make use of the hydrodynamic code
\verb"CLAWPACK"\footnote{\emph{Conservation LAWs PACKage}.  Details
can be found at \texttt{http://www.amath.washington.edu/\~{}claw/}},
an explicit Eulerian, \emph{shock-capturing} and \emph{upwind}
second-order Godunov (with TVD stability) method.

Our investigation is centered on the impact on an impinging shock
during two main stages of collapse of the target halo: turn-around and
virialization, both of which are described in further detail below. In
both cases, we set the redshift of interaction to be $z = 9$, as a
typical value during reionization, and the halo mass to $M=10^8 h^{-1}
M_\odot$, which is chosen both because of the ubiquity of such objects
at $z=9$ (e.g. Scannapieco, Ferrara, \& Madau 2002) and because they
are able to quickly cool by atomic lines, regardless of the evolution
of primordial $H_2$ \citep{HRL97,C00}.

\subsection{Gas distribution in the target halo}

In constructing a model for the structural properties of the target
halo, we neglect self-gravity of the baryons, which is motivated
by the fact that $\Omega_b/\Omega_{\rm m} \ll 1$. Therefore we consider
the gas in hydrostatic equilibrium inside the gravitational potential
of the dark matter halo.

\subsubsection{The virialized case}
\label{virialized_case}

To determine how the baryons are arranged in the dark matter gravitational field, we must first
determine the dark matter distribution inside pre-galactic systems. Here we assume that dark
matter halos at virialization equilibrium follow the universal (spherically averaged) density
profile determined by Navarro, Frenk \& White (1997; hereafter NFW):
\begin{equation}
\label{model:profNFW}
 \rho(r) = \frac{\rho_c \Omega_m (1+z)^3 \, \delta_c}{\displaystyle c x \left( 1 + c x
 \right)^2} \ ,
\end{equation}

\nit where

\begin{displaymath}
x \equiv \frac{r}{R_{\rm vir}} \ ,
\end{displaymath}
$\rho_c$ is the critical density of the universe, $R_{\rm vir}$ the (physical) virial radius of
the system (at which the mean enclosed density is 200 times the mean cosmic value
$\overline{\rho}_m = \rho_c \Omega_m (1+z)^3$), $c$ is the halo concentration parameter, $\delta_c
= (200/3)c^3/F(c)$ the characteristic overdensity, and

\beq
 F(c) \equiv \ln (1 + c) - \frac{c}{1 + c} \ .
\eeq

\nit The total mass of the halo within the virial radius and the mass within a radius $r <
R_{\rm vir}$ are, respectively:

 \beq
 M = \frac{4 \pi}{3} \, 200 \rho_c \Omega_m (1+z)^3 \, R_{\rm vir}^3 \ ,
 \eeq
and
 \beq
 M(r) = \int_0^r \rho(r^{\prime}) \, 4 \pi r^{\prime \, 2} \, \ud r^{\prime} = M \frac{F(c x)}{F(c)} \ .
 \eeq

\nit Equation (\ref{model:profNFW}) allows us to calculate a
circular velocity squared

\beq
 v_c^2(r) = \frac{G M(r)}{r} = V_c^2 \frac{F(c x)}{x F(c)} \ ,
\eeq

\nit where $V_c \equiv (G M / R_{\rm vir})^{1/2}$ is the value at the virial radius.
The work done by the gravitational force on the escaping gas is equal to the variation of its
kinetic energy, therefore the gas at the radius $r$ is able to escape provided it has a velocity
(squared) larger than:

 \beq \label{model:vfugar}
 v_e^2(r) = 2 \int_r^{\infty} \frac{G M(r')}{r'^{\,2}} \, \ud r' = 2 V_c^2 \, \frac{F(cx) + cx/(1+cx)}{x
 F(c)} \ ,
 \eeq
in which the NFW profile is extrapolated to infinity.
The escape speed is maximum at the center of the halo, $v_e^2(0) = 2 V_c^2 c/F(c)$.

To proceed, we follow the algorithm described in the appendix of NFW,
and compute the concentration parameter (or, equally, the
characteristic density contrast $\delta_c$) of dark matter halos as a
function of their mass for our adopted cosmological model.
The algorithm assigns to each halo of mass $M$ identified at redshift
$z$ a collapse redshift $z_{\rm coll}$, defined as the redshift at which half
of the mass of the halo was first contained in progenitors more
massive than some fraction of the final mass \citep{LC94}. The
assumption that the characteristic density of a pre-galactic system is
proportional to the critical density at the corresponding $z_{\rm coll}$
implies:

 \beq
 \delta_c (M, z) \propto \left( \frac{1 + z_{\rm coll}}{1 + z} \right)^3 .
 \eeq

\nit Lower mass halos generally collapse at higher redshift, when the
mean density of the universe is greater: for this reason, from the
previous relation we deduce that low-mass systems, at any given time,
are more centrally concentrated than massive ones. As an example, we
report in the following table the values of $z_{\rm coll}$ and $c$,
obtained for different halos identified at $z=9$.

\begin{center}
 \begin{tabular}{||c|c|c||}
 \hline $M \, [M_\odot/h]$ & $z_{\rm coll}$ & $c$ \\ \hline \hline
 $10^7$ & 12.5 & 4.9 \\ \hline $10^8$ & 12.2 & 4.8 \\ \hline $10^9$ &
 11.9 & 4.7 \\ \hline
 \end{tabular}
 \end{center}

To study the impact of high-redshift cosmic explosions on objects at
$1+z \approx 10$ in detail, we will assume a ``typical" concentration
parameter of $c=4.8$. At these epochs, the dark matter halo of a
subgalactic system is characterized by a virial radius:

\beq \label{model:Rvir} R_{\rm vir} = 0.76 \; M_8^{1/3} \,
 \Omega_m^{-1/3} \, h^{-1} \left( \frac{1 + z}{10} \right)^{-1} \quad
 \textrm{kpc}, \eeq

\nit a circular velocity at virial radius:

\beq V_c = 24 \; M_8^{1/3} \, \Omega_m^{1/6} \left( \frac{1 + z}{10}
 \right)^{1/2}  \quad \textrm{km s}^{-1}, \eeq

\nit and a virial temperature:

\beq \label{model:Tvir} T_{\rm vir} = \frac{G M}{R_{\rm vir}}
 \frac{\mu m_p}{2 k} = 10^{4.5} \; M_8^{2/3} \, \Omega_m^{-1/3} \, \mu
 \left( \frac{1 + z}{10} \right) \quad \textrm{K}, \eeq

\nit where

 \begin{displaymath}
 M_8 \equiv \frac{M}{10^8 \, M_\odot \, h^{-1}},
 \end{displaymath}

\nit and $\mu$ is the mean molecular weight ($\mu = 0.59$ for a fully
ionized hydrogen/helium gas of primordial composition). The escape
speed at the center is:

 \beq v_e (0) = 77 \; M_8^{1/3} \, \Omega_m^{1/6} \left( \frac{1 +
 z}{10} \right)^{1/2}  \quad \textrm{km s}^{-1}.  \eeq

\nit Note, however, that high-resolution $N$-body simulations by
\cite{Buetal01} seem to indicate that high-redshift halos are actually
less concentrated than expected from the NFW prediction. If this is
the case, we might slightly overestimate the values of the escape
speeds from pre-galactic systems.

Having fixed the dark matter density distribution, we can establish
the density profile of the baryons. If the gas collapses and
virializes along with dark matter, it will be shock-heated to the
virial temperature and will settle down to an isothermal profile \citep{MSS98}:
\beq \label{model:profbar} \rho_b (r) = \rho_0 \, e^{ \textstyle - A
\left[ v_e^2(0)  - v_e^2(r) \right]} , \eeq
\nit with
\begin{displaymath}
A = \frac{\mu m_p}{2 k T_{\rm vir}} .
\end{displaymath}

\nit Throughout this paper we make the assumption that the gas is
distributed isothermally in virialized structures.  \nit The central
density of the gas $\rho_0$ is determined by the condition that the
total baryonic mass fraction within the virial radius is equal to
$\Omega_b/\Omega_m$ initially:
\beq \frac{\rho_0}{\rho_c \, (1+z)^3} = \frac{(200/3) \, c^3 \,
 \Omega_b \, e^B} {\int_0^c t^2 \, (1+t)^{B/t} \, \ud t} = 44000 \
 \Omega_b,
\eeq
\nit where $B \equiv 2c / F(c)$. At the virial radius $\rho_b (R_{\rm
vir}) = 0.00144 \, \rho_0$.

\subsubsection{The turn-around case}
\label{turn-around_case}

Once a positive density perturbation has collapsed, it continues to
attract matter.  Bound shells cease expanding with the Hubble flow,
continue to turn-around, and consequently the mass of the perturbation
grows by accretion of new material. This smooth accretion is generally
called \emph{secondary infall} in the literature, because particles
fall onto an already formed object. In this case we make use of the
results obtained by \cite{Be85}. Since gravity has no preferred scale,
we can find self-similar solutions for secondary infall and accretion
onto an overdense perturbation.  The behaviour of this infall depends
on the nature of the gas (collisional or collisionless) and on central
boundary conditions, such as the presence of a black hole in the
core. In our situation, the collisional (baryonic) matter moves in the
potential well generated by the dissipationless (dark-matter)
component. The radius of the shell that is turning around increases
with time as $t^{8/9}$, while the mass within this radius increases as
$t^{2/3}$. At large radii, both dark matter and baryonic
components have the same density distribution, given by: \beq
 \label{model:profTA}
 \rho(r) \propto r^{-9/4} \ .
 \eeq

\nit At smaller radii, the main difference between the two components is the existence of a
spherical shock that forms in the baryonic gas due to the collision of the infalling gas with the
self-generated core.  Since this gas is dissipative, infalling fluid elements are decelerated by
passage through this shock and by post-shock pressure gradients, and come to rest at the
center.

In this model, fluid motion is unchanged when parameters are conveniently scaled as $r \rightarrow
\lambda = r/R_{\rm ta}$:

 \beq
 v(r,t) = \frac{R_{\rm ta}}{t} \, \hat{v}(\lambda)
 \eeq
 \beq
 \rho_b(r,t) = \rho_{\rm igm} \, \hat{\rho}(\lambda)
 \eeq
 \beq
 p(r,t) = \rho_{\rm igm} \left(\frac{R_{\rm ta}}{t}\right)^2 \hat{p}(\lambda)
 \eeq
 \beq \label{model:massTA}
 M_b(r,t) = \frac{4}{3} \pi \rho_{\rm igm} R_{\rm ta}^3 \hat{M}(\lambda)
 \eeq

\nit where $v$, $\rho_b$, $p$, $M_b$ are respectively the gas velocity, density, pressure and the
mass at the distance $r$ from the centre of the perturbation at the cosmological time $t$, while
the same symbols with the ``hat" represent the relative dimensionless quantities; $R_{\rm ta}$ is the
turn-around radius and $\rho_{\rm igm}$ is the mean baryonic density of the intergalactic medium at the
redshift of interest (see equation (\ref{model:densigm})). The shock is located at a fixed
fraction of the turn-around radius, $\lambda_s = 0.347$ for a gas with $\gamma = 5/3$, and
propagates outwards according to the scaling $r_s \propto t^{8/9}$. The total mass of the
over-dense region within the turn-around radius is calculated with $\hat{M}(\lambda=1) = 5.6$
and with $\Omega_m$ instead of $\Omega_b$ in (\ref{model:massTA}):

 \beq
 \label{model:Rta}
 R_{\rm ta} = \left({3 M\over {4\pi} \, 5.6 \, \rho_c \Omega_m}\right)^{1/3} (1 + z)^{-1} .
 \eeq

\subsection{The intergalactic medium}
\label{intergalactic medium}

The intergalactic medium around the simulated halo is considered to be homogeneous with a density
$\rho_{\rm igm}$:

 \beq \label{model:densigm}
 \rho_{\rm igm} = \rho_c \, \Omega_b \, (1 + z)^3 = 1.88 \times 10^{-29} \, h^2 \, \Omega_b \, (1 + z)^3
 \quad \textrm{g cm}^{-3}
 \eeq

\nit corresponding to a numerical density of about $4 \times 10^{-4}$ cm$^{-3}$ at redshift $z=9$.
Furthermore, we assign to the IGM a temperature of $T = 10^4$ K, typical of a fully ionized medium.
This value is explained by the observations of absorbing lines of the IGM in spectra of distant
quasars \citep{Ra98}. It follows that the sound speed is $C_{\rm igm} \simeq 15$ km s$^{- 1}$.

\subsection{The shock wave}

To determine the properties of the impinging shock, we adopt a thin-shell approximation.  In this
model, we require an estimate of the shock Mach number $\mathcal{M}$ and the thickness of the
cooled shell.  If radiative losses are small within the hot-bubble driving the cosmic explosion,
the shock velocity can be approximated by a Sedov-Taylor blast-wave solution:
 \beq \label{model:shockvel}
 v_s = \left(\frac{2 K \epsilon_k E}{3 \pi \rho_{\rm igm}}\right)^{1/2} d^{-3/2} ,
 \eeq

\nit where $E$ is the total energy of the shock, $\epsilon_k = 0.2$ (e.g.\ Scannapieco, Ferrara \&
Madau 2002) is the total-to-kinetic energy conversion efficiency, and $K = 1.53$ is a constant
obtained from the exact solution for $\gamma = 5/3$. As for the energy $E$, we assume that one
supernova occurs for every $100 \, M_\odot$ of baryons that went into stars (see, e.g.,
\cite{Gi97} 1997), so the number $N$ of supernovae driving the bubble is:

\begin{displaymath}
N = 5 \times 10^4 M_8 \, \epsilon_{\rm sf} \, (\Omega_m \, h)^{-1} ,
\end{displaymath}

\nit where $\epsilon_{\rm sf}$ is the initial star formation efficiency. The total energy of the
shock is then:

 \beq
 \label{model:totenergy}
 E = 2\times 10^{51} \epsilon_k N  \quad \textrm{erg},
 \eeq

\nit where $2 \times 10^{51}$ erg is the energy of each supernova, taking into account also the
contribution from stellar winds. Assuming an initial star formation efficiency $\epsilon_{\rm sf} =
0.1$, a value that is broadly supported by observations of high-$z$ IGM metals and the high-$z$
cosmic star formation rates \citep{C00,Bark02,SFM02}, we can estimate $N \simeq 5000 \,
\Omega_m^{-1}$ and thus $E \simeq  10^{55} \epsilon_k \Omega_m^{-1}$ erg for typical dwarf
galaxy-sized objects of mass $M \sim 10^8 M_\odot$.

We estimate the distance $d$ in equation (\ref{model:shockvel}), i.e. the radius of the shock
wave, as the mean spacing between halos of a mass scale $M$:

 \beq d \approx \left({3 M\over {4\pi} \rho_c \Omega_m}\right)^{1/3} (1 + z)^{-1} = 94 \, M_8^{1/3} (1
 + z)^{-1} \ {\rm kpc} . \label{model:meansep}
 \eeq

\nit From equations (\ref{model:densigm}), (\ref{model:totenergy}) and (\ref{model:meansep}) we
note that the value of $v_s$ in equation (\ref{model:shockvel}) is independent of redshift and halo
mass.

An alternative estimate relies on the fact that the mean number of halos within a distance $r$
from another halo is:

 \beq \Psi(\leq r) = 4 \pi n  \int_{R_{\rm vir}}^r r^{\prime\,2} \, [1+\xi(r')] \, \ud r' \, ,
 \label{eq:nr}
 \eeq

\nit where $n$ is the average number density of halos and $\xi$ is their two-point correlation
function.  Both these quantities can be estimated from the classic analytical peaks formalism
(Press \& Schechter 1976, hereafter PS; Bond \etal 1991; Mo \& White 1996). Note that the average
number density in this equation is far less than $(\frac{4\pi}{3} d^3)^{-1}$ as defined by equation
(\ref{model:meansep}). This is due to the fact that the mean separation between halos is much
larger than the average distance between a halo and its closest neighbor, because they are
clustered together into groups of objects that only occupy a fraction of the cosmological volume.
From equation (\ref{eq:nr}), we can compute the probability for a nearest neighbour to be at a
distance $r$ as $P(r) = 1-\exp[-\Psi(\leq r)]$ and the mean distance to the nearest neighbour as:

 \beq
 d = \int_{R_{\rm vir}}^\infty r \frac{\ud P}{\ud r}(r) \, \ud r \, .
 \eeq

\nit In this case, approximating the physical density $n$ as $(1+z)^{-3}$ times the PS number
density of $\frac{\ud n}{\ud \ln{M}}(M = 10^8 \msun/h, \, z = 9)$, and making use of the matter
power spectrum from Eisenstein \& Hu (1998), with $\sigma_8 = 0.87$, we obtain a distance of $14$
kpc.  The similarity of this (power-spectrum and redshift dependent) estimate with our more
straightforward approach, give us confidence that the distance $9.4$ kpc as given by equation
(\ref{model:meansep}) is a reasonable estimate for interactions in which we are interested.

We have to assess now the value of the second parameter of interest for the shock wave, the
thickness of the shell $\Delta d$. Almost the totality of the gas swept up by the front is
concentrated in a relatively thin layer and the density inside the layer is constant and equal to
the density behind the front surface. We have also to consider the fraction of baryons removed
from the outflowing galaxy: according to \cite{MFM02}, we can take about the $60\%$ of a system
with the same mass scale $M$ of interest. In conclusion we obtain for the thickness $\Delta d$ of
the shell, from conservation of mass:

 \beq \label{model:shock_thickness}
 \Delta d = d \, \frac{\rho_{\rm igm}}{3 \, \rho_s} + \frac{0.6 \, M \, \Omega_b}{4 \pi \, \rho_s \, d^2} ,
 \eeq

\nit where $\rho_s$ represents the post-shock density (i.e. the density inside the layer), achieved
with the usual well-known``jump conditions" ($\rho_s \simeq 4 \, \rho_{\rm igm}$ for supersonic
shocks).

\section{Simulation results}
\label{simul}

In both virialized and pre-virialized cases, we consider a computational grid made of $150 \times
150$ cells and set the boundary conditions as zero-order extrapolation, i.e. requiring that the
flow of the matter runs only outward across all four borders.  As mentioned above, in both cases
we select a target halo mass of $M=10^8 h^{-1} M_\odot$ and redshift of $z= 9$ as our fiducial
model.
Convergence tests with double spatial resolution of the grid ($300 \times 300$ cells) have
been made to ensure that the simulations provide reliable results: in all cases the answers are
consistent.

\subsection{Virialized case}
\nit Let us first consider the virialized case. For our fiducial parameters the virial radius of
the pre-galactic system, which is assumed to be spherically symmetric, is equal to $1.75$
physical kpc and
its virial temperature is $T_{\rm vir} = 12,490$ K. In this case,
the spatial resolution of the computational grid
is $43.3$ pc. To determine the velocity at which the shock reaches the halo, we replace the
distance $d$, in eq.\ (\ref{model:meansep}) with:

 \beq \label{simul:new_distance}
 d \rightarrow d' = d - R_{\rm vir} ,
 \eeq

\nit as we are interested in the initial point of the collision, when the front first reaches the
overdense region.

\nit The simulation starts with the coordinates of the centre of the target halo placed at the
point\footnote{Hereafter we will indicate the coordinate of the cell along the horizontal
axis as $x$ and the coordinate along the vertical axis as $y$.}
$(x_0,y_0) = (63,75)$, i.e. $(2.7,3.2)$
kpc, with a radius of $\sim 40$ cells. Both the pre-galactic system and the intergalactic medium
are at rest, and the shock wave is modeled as a plane wave,
at $x = 20$ cells, moving from the left towards the right, with an initial velocity of about $214$
km s$^{-1}$. In this case, the numerical density $n_s$ and the pressure $p_s$ behind the shock are
$n_s = 4n_{\rm igm} = 1.6 \times 10^{-3}$ cm$^{-3}$ and $p_s \simeq \rho_{\rm igm} \, v_s^2,$
respectively, and the temperature of the post-shock gas reaches a value of $6.2 \times 10^5$ K.

Let $t = 0$ be the time referred to by the initial conditions. The final time of the simulation is
then $t = 58.2$ Myr, equal to about $3.6$ times the shock crossing time of the halo, defined as:

 \beq \label{simul:tcross}
 t_{\rm cross} = \frac{2 R_{\rm vir}}{v_s} \ .
 \eeq

\nit Note that virialization is quite late in the evolution of the perturbation, and during our
simulation time it is possible that star formation (and SNe) could occur inside this object.
While, in this case, the dynamics of the system would become more complicated, we choose to ignore
these effects and study the virialized case as an idealization of the {\em last possible time} at
which baryonic stripping could prevent star formation in  the collapsing object. Indeed, due to
the short time between virialization and star formation, it is much more likely that baryonic
stripping will occur prior to virialization, as we discuss in more detail below.

 \begin{figure*}
\includegraphics[width=\textwidth,height=21cm,width=14cm]{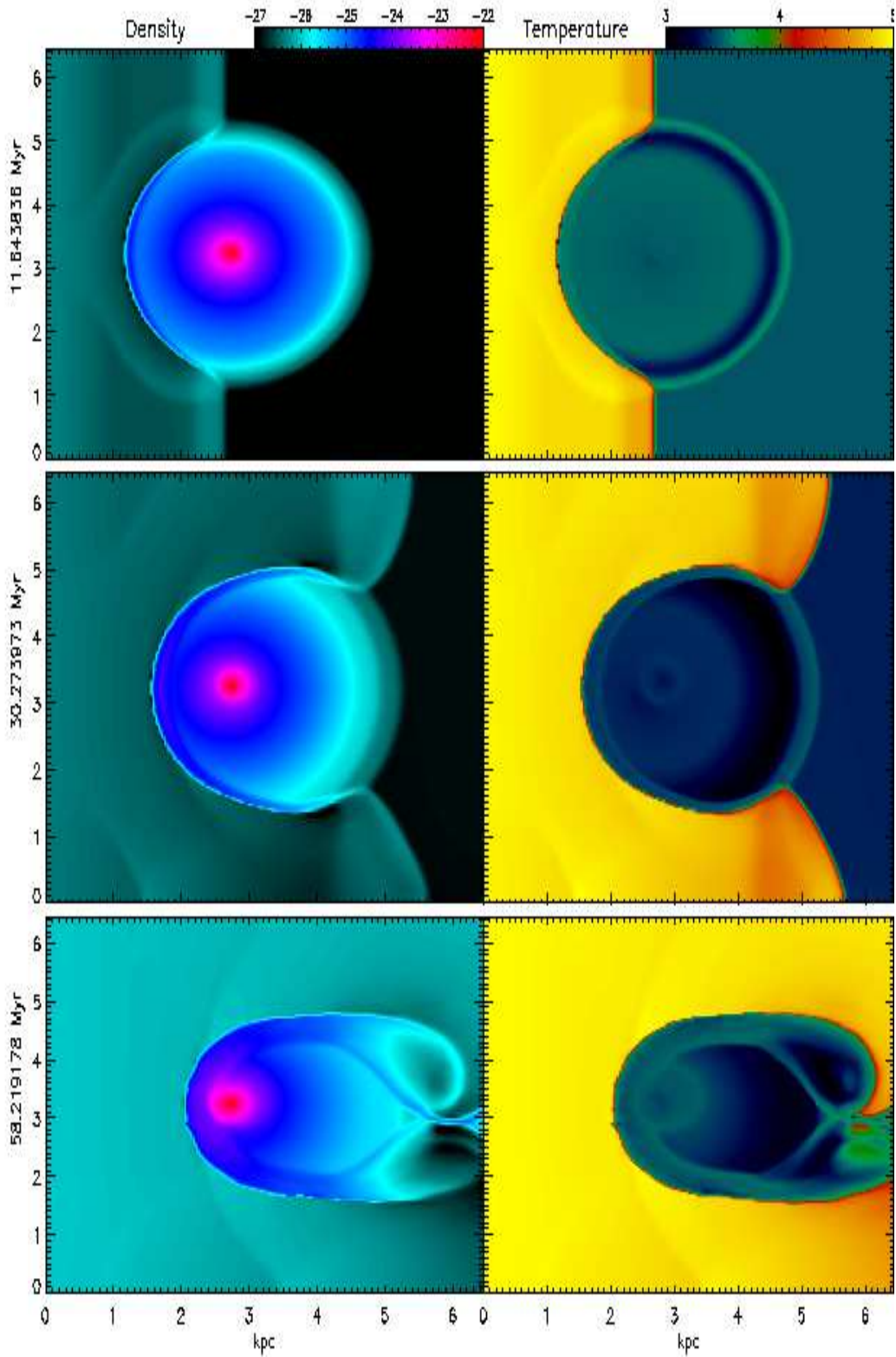}
\caption{Snapshots of the dynamical evolution for the virialized halo case (NFW profile).
Maps with logarithmic color scale are plotted for the density [g cm$^{-3}$] (left)
and temperature [K] (right) fields at three different times ($t = 11.6, \ 30.3, \ 58.2$ Myr).}
\label{fig:maps_virial}
 \end{figure*}

In Figure~\ref{fig:maps_virial} we show the density and temperature fields at three key stages.
Initially the shock wave travels towards the centre, compressing the gas. When the shock first
reaches the system, it encounters a density slightly greater than 30 times that of the IGM. At
this point, the velocity of the internal wave changes in direction and drops by a factor of
approximately the square root of ratio of the halo $\rho_b$ and IGM $\rho_{\rm igm}$ densities
\citep{BP75}:

 \beq \label{simul:velshockalone}
 v_{s, \, {\rm h}} \approx \frac{v_{s, \, {\rm igm}}}{( \rho_b/\rho_{\rm igm} )^{1/2}} \ ,
 \eeq

\nit where $v_{s, \, {\rm h}}$ and $v_{s, \, {\rm igm}}$ are the front velocities inside the halo
and in the IGM respectively, which gives $\sim 38$ km s$^{-1}$.

During the second stage, the shock velocity gradually decreases,
according to equation (\ref{model:profbar}) as it runs into denser
regions. In particular, the front reaches $\mathcal{M} \sim 1$
close to the halo core and lapses into a sonic wave, mingling with the
surrounding matter. At this time, the shock begins to sag and to form
two symmetric extensions in which the baryons are accumulated.

In the meantime the outflow continues to propagate outside the halo,
with a velocity greater than the velocity inside the cloud. The
intergalactic gas has now a post-shock pressure of $\sim 1.4 \times
10^{-13}$ erg cm$^{-3}$, more than $240$ times the IGM pressure and
about $6$ times the baryonic pressure of the collapsed system at the
virial radius.

Finally, at a time $t \simeq 50$ Myr the two flux sides join together
again behind the halo, ``embracing'' the entire system. After this
time it is unlikely that the mass inside the virial radius will be
ejected by momentum transfer. The dynamics of the interaction
then becomes quite complicated, due to the presence of
reflections of shock waves which scatter in  the perturbation.

\begin{figure*}
\includegraphics[width=14cm]{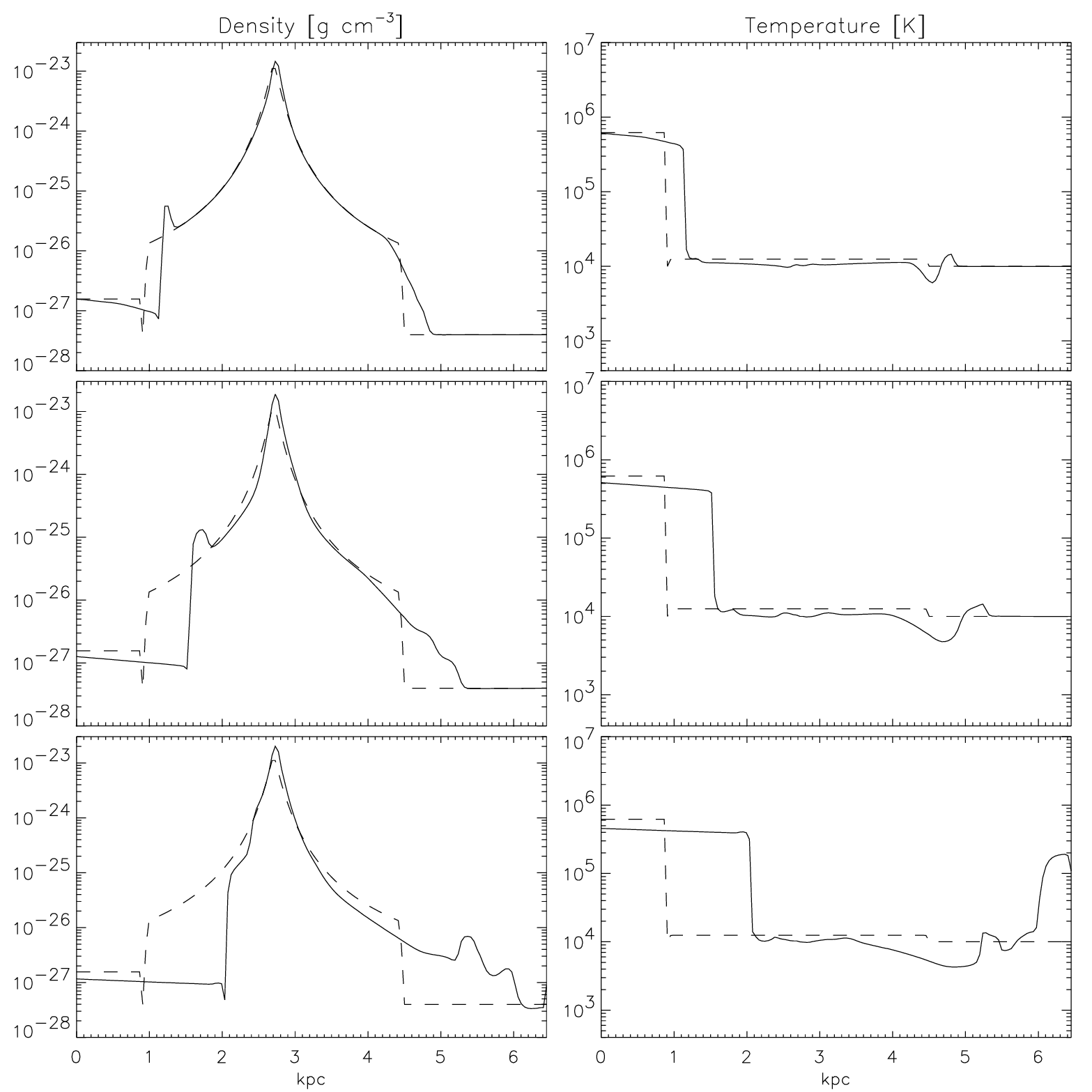}
\caption{Virialized case.
Density (left) and temperature (right) profiles for a horizontal cut crossing the centre of the
halo ($y = 75$), at times (from top to bottom) $t = 11.6, \ 30.3, \ 58.2$ Myr. Dashed lines show
the profiles for the same cut at the initial time.} \label{fig:tagli_virial}
\end{figure*}

In Figure~\ref{fig:tagli_virial} we show density and temperature profiles along the horizontal
direction crossing the halo centre ($y = 75$), at each of these three different stages (from top
to bottom).
These plots uncover the same trends Figure~\ref{fig:maps_virial}, and indicate that after the
interaction the gas structure of the halo is only weakly modified, especially in the core, where
it is nearly unchanged.

 \begin{figure*}
\includegraphics[width=\textwidth,height=21cm,width=14cm]{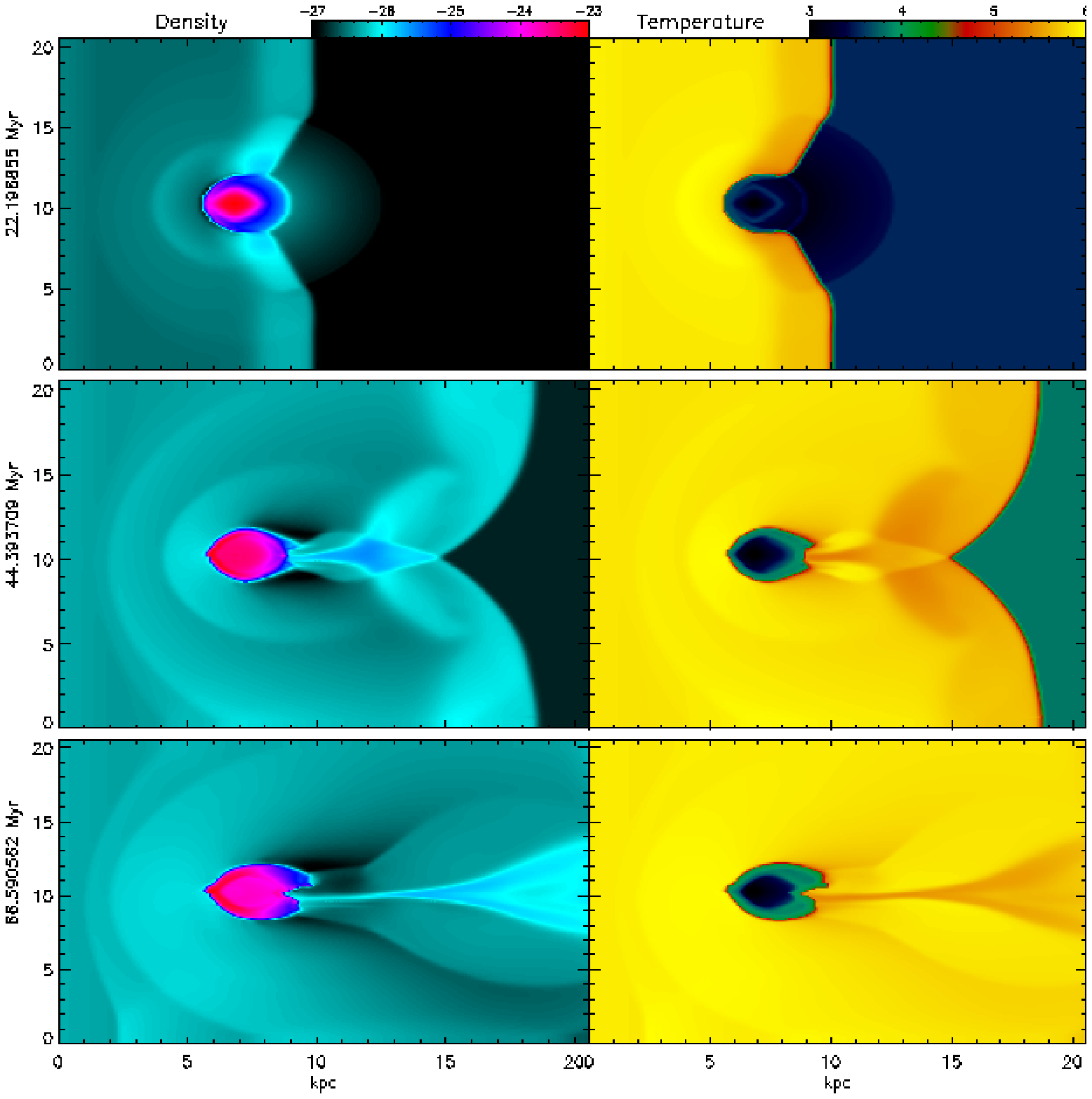}
\caption{Snapshots of the dynamical evolution for the halo at the turn-around stage.
Maps with logarithmic color scale are plotted for the density [g cm$^{-3}$] (left)
and temperature [K] (right) fields at three different times ($t = 22.2, \ 44.4, \ 66.6$ Myr).}
\label{fig:maps_turnaround}
 \end{figure*}

%\clearpage

\begin{figure*}
\includegraphics[width=14cm]{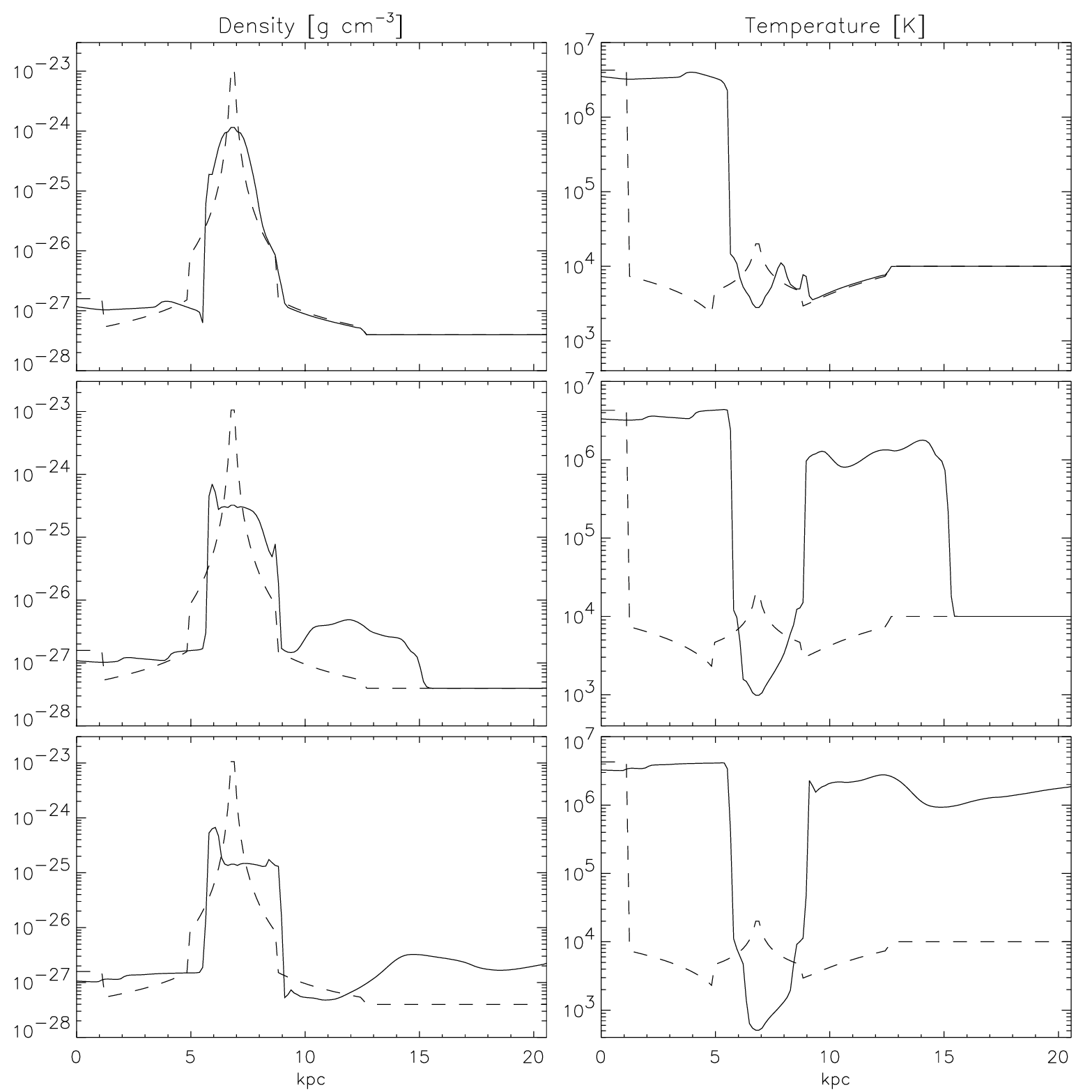}
\caption{Turn-around stage. Density (left) and temperature (right) profiles for a horizontal cut
crossing the centre of the halo ($y = 75$), at times (from top to bottom) $t = 22.2, \ 44.4, \
66.6$ Myr. Dashed lines show the profiles for the same cut at the initial time.}
\label{fig:tagli_turnaround}
 \end{figure*}

\subsection{Pre-virialized case}

Let us move now to the pre-virialized case, when the halo is at the
turn-around stage. Here the radius of our fiducial system, with the
same mass and at the same redshift, is given by equation
(\ref{model:Rta}), which yields $R_{\rm ta} = 5.7$ kpc. Therefore, the
Mach number of the shock  wave given by equation
(\ref{model:shockvel}) is larger than the previous value, and  we
obtain $\mathcal{M} = 37$ (i.e. a front expanding in the IGM with an
initial velocity of roughly $565$ km s$^{-1}$ and with a crossing time
of the halo  $t_{\rm cross} \simeq 19.8$ Myr) and $\Delta d \simeq
1.2$ kpc. The post-shock temperature of the gas reaches the value of
$T \sim 4.3 \times 10^6$ K. The new configuration of this problem
consists of a spatial resolution of about $138$ pc in the same $150
\times 150$ computational grid. The centre of the overdense region is
put at $(49,75)$ and its radius encloses $41$ cells. The final time of
the simulation is $133$ Myr.

The dynamics (Figure~\ref{fig:maps_turnaround}) of this interaction,
by and large,  follow the same sequence discussed in the virialized
case. Nevertheless some important differences can be seen.  Note that
at this stage the halo is not yet isothermal and the solutions by
\cite{Be85} expect values of pressure at the edge almost equal to
those of the IGM, and a smoother decrease of the density with radius.

It follows that at this epoch the shock will be able to more easily
penetrate into the centre of the
perturbation. Figure~\ref{fig:maps_turnaround} points out this
fragility, showing that now the front is able to reach the halo
core. For further details see also Figure~\ref{fig:tagli_turnaround},
representing the evolution in time (from top to bottom) of the density
and temperature profiles along the horizontal cut crossing the halo
centre ($y = 75$). The comparison between Figure~\ref{fig:maps_virial}
and Figure~\ref{fig:maps_turnaround} emphasizes a morphological
difference of the system.

\section{Quantitative analysis}
\label{results}

As the primary aim of our investigation is to assess the impact of
galactic outflows on neighbouring overdense regions, it is important
to quantify the amount of matter that can be removed from the halo. We
follow the two criteria proposed by \cite{SFB00} to estimate this
quantity. There are two mechanisms by which outflows from nearby
objects can hinder the formation of a galaxy: in one case the gas is
heated to a temperature greater than the virial temperature
(``mechanical evaporation"), in the other case the shock transfers
sufficient momentum to carry gas parcels out of the gravitational
potential (``baryonic stripping").

\begin{figure}
\includegraphics[width=7.7cm]{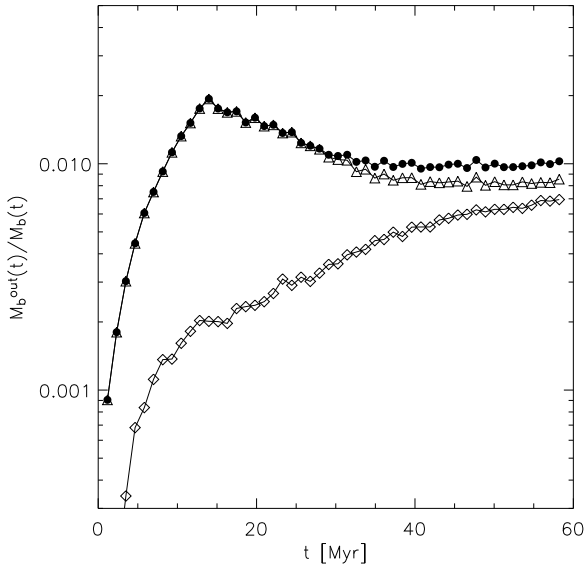}
\caption{Mass loss in the virialized case. Triangles represent the fraction of mass with $T
\geq
T_{\rm vir}$, while the diamonds points give the fraction of mass with $v \geq v_e$. Line with filled
circles gives the total fraction of the baryonic mass removed from the virial radius of the halo by
both mechanisms.} \label{fig:massout_virial}
\end{figure}

\begin{figure}
\includegraphics[width=7.7cm]{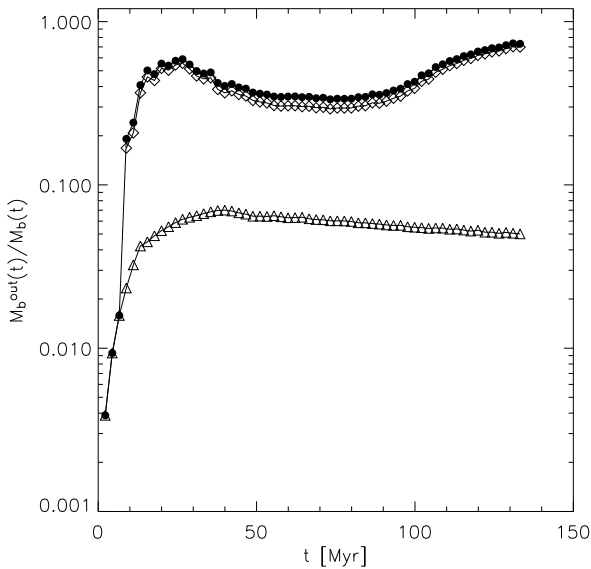}
\caption{Mass loss in the halo at turn-around. Lines and points are as in Figure
\protect\ref{fig:massout_virial}.  Note the difference in scale on the $y$-axis.}
\label{fig:massout_turnaround}
\end{figure}

Figure~\ref{fig:massout_virial} shows, for the virialized case, the total fraction of baryonic
mass removed from the virial radius of the halo (filled circles), by each of these mechanisms. The
triangles (diamonds) represent the evolution of the fraction of mass with $T \geq T_{\rm vir}$,
($v(r) \geq v_e(r)$), integrated over the halo. The first condition describes the importance of
the mechanical evaporation, whereas the second one refers to baryonic stripping.

The physical interpretation of the behavior of the curves is as follows.  The initial growth of
the lines is due to the penetration of the shock into the outer layers of the halo, which are much
less dense than the central core. Even at these large radii, however, the escape velocity is
nearly 38 km s$^{-1}$  (see the lower curve), i.e. $v_e \geq v_s$
(\ref{simul:velshockalone}). Thus only a tiny fraction of the virialized gas
is susceptible to baryonic stripping.

As for the thermal contribution, while the gas on the outskirts of the
perturbation is heated above $T_{\rm vir}$, causing the initial rise
in the upper curve, the short cooling times in the dense inner regions
of the halo prevent $T \geq T_{\rm vir}$ heating at later times. A
heating \emph{plateau} is soon achieved, after which additional
baryonic heating becomes negligible. Thus at the end of the simulation
($t=58.2$ Myr), only  $\sim 0.9 \%$ of the gas is affected by
mechanical evaporation and only $\sim 0.7 \%$ is supplied with
 enough
momentum to be stripped; hence, the impact of the cosmic blast is
trifling.

Figure~\ref{fig:massout_turnaround} shows the analogous plots for the
pre-virialized halo at the turn-around phase.  In this case we find
that galaxy formation is strongly affected, as a result of the
stronger shock (the Mach number is larger than in the previous case)
and the more diffuse structure of the over-dense region, which make
the system more vulnerable to the impinging blast.  This is
particularly evident in the case of baryonic stripping, which
increases remarkably ($\sim 69.9 \%$ at the end of the simulation,
$t=133$ Myr) in comparison with the virialized case. On the other hand
the behaviour of the thermal contribution (curve with triangles) is
only weakly enhanced with respect to the previous case, attaining a
value of  $\sim 5.0 \%$ at the final output time.

\section{Summary and Discussion}
\label{summary}

In this paper we have studied the interaction of a cosmic shock wave,
generated by multiple supernovae occurring in a primordial galaxy,
with a neighbouring halo either already virialized or at the
turn-around stage. The main aim of the investigation is to assess if
an intergalactic shock may be able to prevent the collapse of a
fraction of dwarf galaxies at high redshifts, thus alleviating the
so-called ``cooling catastrophe'' (for a review of feedback effects
see \cite{FS04}). We have followed in detail the evolution of the
target halo during and after the impact with the supersonic front by
means of numerical simulations.

The main results of this study can be summarized as
follows. Suppression of dwarf galaxy formation can occur in the
collapse stage at or before turn-around.  In this case we find that
about 70\% of the gas can be stripped from the parent halo and ejected
into the intergalactic medium. Most of the gas is unbound by the
impinging momentum  from the shock wave, which is sufficient to
accelerate it to velocities larger than the local escape speed. The
effects of thermal evaporation of the gas (i.e. due to gas heating by
the shock) are found to be negligible in comparison to stripping, in
agreement with the conclusions of \cite{SFB00}.

The effect of the shock wave is much weaker if the halo is already in
a virialized state because of the more compact configuration and
consequent higher gas density, which enhances radiative losses.  The
shock removes $\approx 1\%$ of the gas, and the forming galaxy is
hardly affected by the interaction.

As the time spent around turn-around is much longer than the
interval from virialization to star formation due to strong
radiative losses, we expect the pre-virialized case to be more
representative of the majority of outflow-galaxy interactions. Thus
SN-driven pregalactic outflows may be an efficient mechanism for
inhibiting the formation of neighbouring galaxies at high redshift.

Our numerical experiments have explored two typical situations among
the possible shock - target galaxy interactions. However, they are by
no means exhaustive. For example we have used mean values in order to
choose the target galaxy (mass, formation redshift, density profile)
and shock properties (energy, velocity at the interaction). Therefore
a full exploration of the likely range for these parameters should be
considered, at least for the more interesting case of pre-virialized
halos, before a final conclusion on the relevance of these event can
be drawn.  Additional complications may arise from occurrences of
multiple shocks impinging on the same halo, coming from different
source galaxies. However, given the relatively short duration of the
interaction with the shock (100-200 Myr) the probability that more
than one shock at a time is acting on the target galaxy seems
unlikely, as this would require remarkable synchronization of the
starbursts in different galaxies.

The implications of the suppression effects might be particularly evident in the shape and
evolution of the luminosity function. In particular, we expect a selective decrease in the number
of low-luminosity objects, due to their inability to collapse and form stars. Our results might
also have implications for the number of satellites found around large galaxies, one of the major
problems in current hierarchical models of galaxy formation. The baryonic component of these
satellites could be destroyed early on by interactions with shocks produced by the first cosmic
episodes of star formation. Finally, stripping might drastically change the baryonic-to-dark
matter ratios as a function of galactic mass. Although it is likely that SN explosions inside a
given galaxy will affect this observable, our study shows that impinging intergalactic shocks can
play an equally important role. We plan to investigate these aspects in forthcoming work.

\section*{Acknowledgments}

We are grateful to Chris McKee for helpful comments during the preparation of this manuscript.
We would also like to thank an anonymous referee for useful discussion. ES was supported
by an NSF Math and Physical Sciences Distinguished International Postdoctoral Research (NFS
MPS-DRF) fellowship during part of this investigation; his research was also supported by the
National Science Foundation under grant PHY99-07949. We acknowledge partial support from the
Research and Training Network ``The Physics of the Intergalactic Medium" established by the
European Community under the contract HPRN-CT2000-00126 RG29185.


\begin{thebibliography}{}

\bibitem[Barkana(2002)]{Bark02}
Barkana, R. 2002, New A., 7, 85

\bibitem[Benson et al.(2001)]{Beetal01}
Benson, A.~J., Lacey, C.~G., Frenk, C.~S., Cole, S., \& Baugh, C.~M.\ 2001, Bulletin of the
American Astronomical Society, 33, 871

\bibitem[Benson et al.(2002a)]{Beetal02a}
Benson, A.~J., Lacey, C.~G., Baugh, C.~M., Cole, S., \& Frenk, C.~S.\ 2002a, MNRAS, 333, 156

\bibitem[Benson et al.(2002b)]{Beetal02b}
Benson, A.~J., Frenk, C.~S., Lacey, C.~G., Baugh, C.~M., \& Cole, S.\ 2002b, MNRAS, 333, 177

\bibitem[Benson et al.(2003)]{Beetal03}
Benson, A.~J., Frenk, C.~S., Baugh, C.~M., Cole, S., \& Lacey, C.~G.\ 2003, MNRAS, 343, 679

\bibitem[Benson \& Madau(2003)]{BM03}
Benson, A.~J.~\& Madau, P.\ 2003, MNRAS, 344, 835

\bibitem[Bertschinger(1985)]{Be85}
Bertschinger, E.\ 1985, ApJS, 58, 39

\bibitem[Bond et al.(1991)]{Bond91}
Bond, J. R., Cole, S., Efstathiou, G., \& Kaiser, N. 1991, ApJ, 379, 440

\bibitem[Bullock et al.(2001)]{Buetal01}
Bullock, J.~S., Kolatt, T.~S., Sigad, Y., Somerville, R.~S., Kravtsov, A.~V., Klypin, A.~A.,
Primack, J.~R. \& Dekel, A.\ 2001, MNRAS, 321, 559

\bibitem[Bychkov \& Pikelner(1975)]{BP75}
Bychkov, K.~V.~\& Pikelner, S.~B.\ 1975, Soviet Astronomy Letters, 1, 14

\bibitem[Cen \& Bryan(2001)]{CB01}
Cen, R.~\& Bryan, G.~L.\ 2001, ApJ, 546, L81

\bibitem[\protect\citeauthoryear{Cen \& Ostriker}{1999}]{CO99}
Cen, R. \& Ostriker, J. P. 1999, ApJ, 519, L109

\bibitem[Ciardi et al.(2000)]{C00}
Ciardi, B., Ferrara, A., Governato, F. \& Jenkins, A. 2000, MNRAS, 314, 611

\bibitem[Ciardi, Ferrara \& White(2003)]{CFW03}
Ciardi, B., Ferrara, A. \& White, S. D. M. 2003, MNRAS, 344, L7

\bibitem[\protect\citeauthoryear{Dekel \& Silk}{1986}]{DS86}
Dekel, A. \& Silk, J. 1986, ApJ, 303, 39

\bibitem[Edge \& Stewart(1991)]{ES91}
Edge, A.~C.~\& Stewart, G.~C.\ 1991, MNRAS, 252, 414

\bibitem[Eisenstein \&  Hu (1999)]{EH99}
Eisenstein, D. \& Hu, W. 1999, ApJ, 511, 5

\bibitem[Efstathiou(1992)]{Ef92}
Efstathiou, G.\ 1992, MNRAS, 256, 43P

\bibitem[Ferrara \& Salvaterra(2004)]{FS04}
Ferrara, A. \& Salvaterra, R. 2004, \emph{astro-ph/0406554}

\bibitem[Gibson()]{Gi97}
Gibson, B.~K.\ 1997, MNRAS, 290, 471

\bibitem[Gnedin(2000)]{Gn00}
Gnedin, N.~Y.\ 2000, ApJ, 542, 535

\bibitem[Keres \etal (2004)]{keres04}
Kere\v{s}, D., Katz, N., Weinberg, D. H. \& Dav\'e, R. 2004, MNRAS, submitted, astro-ph/0407095

\bibitem[Haiman, Rees, \& Loeb(1997)]{HRL97}
Haiman, Z., Rees, M., \& Loeb, A. 1997, ApJ, 476, 458 (erratum 484, 985)

\bibitem[Heger \& Woosley(2002)]{HW02}
Heger, A. \& Woosley, S. E. 2002, ApJ, 567, 532

\bibitem[Jones \& Forman(1984)]{JF84}
Jones, C.~\& Forman, W.\ 1984, ApJ, 276, 38

\bibitem[Klein, McKee \& Coella(1994)]{KMC94}
Klein, R. I., McKee, C. F. \& Coella, P. 1994, ApJ, 420, 213

\bibitem[Kogut et al.(2003)]{Koetal03}
Kogut, A.~et al.\ 2003, ApJS, 148, 161

\bibitem[Lacey \& Cole(1994)]{LC94}
Lacey, C.~\& Cole, S.\ 1994, MNRAS, 271, 676

\bibitem[\protect\citeauthoryear{Larson}{1974}]{La74}
Larson, R. B. 1974, MNRAS, 169, 229

\bibitem[Mac Low \& Ferrara(1999)]{MacLF99}
Mac Low, M.~\& Ferrara, A.\ 1999, ApJ, 513, 142

\bibitem[Madau, Ferrara \& Rees(2001)]{MFR01}
Madau, P., Ferrara, A. \& Rees, M.~J.\ 2001, ApJ, 555, 92

\bibitem[\protect\citeauthoryear{Madau \& Rees}{2001}]{MR01}
Madau, P. \& Rees, M. J. 2001, ApJ, 551, L27

\bibitem[Makino, Sasaki \& Suto(1998)]{MSS98}
Makino, N., Sasaki, S. \& Suto, Y.\ 1998, ApJ, 497, 555

\bibitem[Mo \& White(1996)]{Mo96}
Mo, H. J. \& White, S. D. M.  1996, MNRAS, 282, 348

\bibitem[Mo \& White(2002)]{Mo02}
Mo, H. J. \& White, S. D. M.  2002, MNRAS, 336, 112

\bibitem[Mori, Ferrara \& Madau(2002)]{MFM02}
Mori, M., Ferrara, A. \& Madau, P.\ 2002, ApJ, 571, 40

\bibitem[Navarro, Frenk \& White(1997)]{NFW97}
Navarro, J.~F., Frenk, C.~S. \& White, S.~D.~M.\ 1997, ApJ, 490, 493

\bibitem[Navarro \& Steinmetz(1997)]{NS97}
Navarro, J.~F.~\& Steinmetz, M.\ 1997, ApJ, 478, 13

\bibitem[Pettini \& Pagel(2004)]{PP04}
Pettini, M.~\& Pagel, B.~E.~J.\ 2004, MNRAS, 348, L59

\bibitem[Press \& Schechter (1974)]{PS74}
Press, W. H. \& Schechter, P. 1974, ApJ, 187, 425

\bibitem[Rauch(1998)]{Ra98}
Rauch, M.\ 1998, ARA\&A, 36, 267

\bibitem[Scannapieco, Ferrara \& Broadhurst(2000)]{SFB00}
Scannapieco, E., Ferrara, A. \& Broadhurst, T.\ 2000, ApJ, 536, L11

\bibitem[Scannapieco \& Broadhurst(2001)]{SB01}
Scannapieco, E., \& Broadhurst, T.\ 2001, ApJ, 549, 28

\bibitem[Scannapieco, Ferrara \& Madau(2002)]{SFM02}
Scannapieco, E., Ferrara, A. \& Madau, P.\ 2002, ApJ, 574, 590

\bibitem[Songaila(2001)]{So01}
Songaila, A.\ 2001, ApJ, 561, L153

\bibitem[Springel \& Hernquist(2003)]{Sp03}
Springel, V. \& Hernquist, L. 2003, MNRAS, 339, 289

\bibitem[\protect\citeauthoryear{Tegmark, Silk \& Evrard}{1993}]{TSE93}
Tegmark, M., Silk, J. \& Evrard, A. 1993, ApJ, 417, 54

\bibitem[Thacker, Scannapieco \&  Davis(2002)]{TSD02}
Thacker, R. J., Scannapieco, E., \& Davis, M. 2002, ApJ, 581, 836

\bibitem[Theuns, Mo \& Schaye(2001)]{TMS01}
Theuns, T., Mo, H.~J. \& Schaye, J.\ 2001, MNRAS, 321, 450

\bibitem[Thoul \& Weinberg(1996)]{TW96}
Thoul, A.~A.~\& Weinberg, D.~H.\ 1996, ApJ, 465, 608

\bibitem[Voit(1996)]{Vo96}
Voit, G.~M.\ 1996, ApJ, 465, 548

\bibitem[White \& Rees(1978)]{WR78}
White, S.~D.~M.~\& Rees, M.~J.\ 1978, MNRAS, 183, 341



%\bibitem[\protect\citeauthoryear{Tozzi et al.}{2000}]{TMMR00}
%Tozzi, P., Madau, P., Meiksen, A., \&  Rees, M. J. 2000, ApJ, {  528}, 597

\end{thebibliography}
\end{document}